\documentstyle[12pt,aasms4]{article}
\lefthead{SENGUPTA \& KRISHAN}
\righthead{POLARIZATION FROM BROWN DWARF}
\begin{document}

\title{PROBING DUST IN THE ATMOSPHERE OF BROWN DWARFS THROUGH POLARIZATION} 
\author{SUJAN SENGUPTA\footnote{sujan@iiap.ernet.in} AND VINOD KRISHAN }
\affil{ Indian Institute of Astrophysics, Koramangala, Bangalore 560 034,
India }

\begin{abstract}

 Theoretical analysis and observational evidences indicate that
a brown dwarf with effective temperature greater than 1300 K would
have dust cloud in its atmosphere. In this letter, we show
that dust scattering should yield  polarized continuum radiation
from the relatively  warm brown dwarfs and the polarized flux profile
could be a potential diagnosis tool for the optical and the physical
properties of dust grains. The degree of polarization due to multiple
scattering will be more in the optical region if the particle size
is small while significant polarization should be detected in the
infra-red region if the particle size is large.  It is pointed out
that the departure from sphericity in the shape of the object due to
rapid rotation and due to tidal effect  by the companion in a binary
system ensures the disc integrated polarization to be non-zero.

\end{abstract}


\section{INTRODUCTION}

 The synthetic continuum spectra  of the  brown dwarf Gliese 229B has been obtained
by several authors.
Marley et al. (1996), Griffith, Yelle \& Marley (1998), Saumon et al (2000),
obtained an overall good fit of the observed spectra of Gliese 229B over a wide
range of wavelengths and attributed the very rapid decline of the observed
continuum flux in the optical region to the dust particulates in its atmosphere.
On the other hand, Tinney et al. (1998); Kirkpatrick et al. (1999);
 Tsuji, Ohnaka, \& Aoki (1999), Burrows, Marley, \& Sharp
(2000) argued 
that the pressure broadened red wing of the 0.77 $\mu m$ K I doublet and
the 0.5891, 0.5897 $\mu m$ Na I D
could account for the observed features of the continuum flux shortward
of 1.1 $\mu m$ (for a recent review, see Burrows, Hubbard, Lunine \& Liebert 2001). 
The first optical spectrum of a T Dwarf SDSS 1624+0029 favors
the latter explanation(Liebert et al. 2000) and at present
there is enough evidence that dust particulates cannot exist in the 
visible atmospheres of comparatively cool T dwarf such as Gl 229B. However,
dust can be present at the visible height of the atmosphere of
warmer brown dwarfs and L dwarfs with effective temperature
greater than 1300 K.

Very recently, Allard et al. (2001) have presented a detail theoretical investigation
on the effects of dust in the atmospheres of brown dwarfs by considering two
limiting cases: (a) inefficient gravitational settling of the dust grains
and (b) efficient gravitational settling. In the former case the dust is distributed
according to the chemical equilibrium predictions and provides the maximum impact
upon the atmosphere while in the latter case the dust forms and depletes refractory
elements from the gas yielding a minimal effect on the brown dwarf atmosphere.  
When the effective temperature of the object is bellow 1300 K complete 
gravitational settling of dust grains should occur whereas gravitational
settling cannot occur when the effective temperature is above 1800 K.
In the intermediate case ($T_{eff}=1400$ K) an optically thick dust cloud
forms in the visible atmosphere (Marley and Ackerman 2001).

In this letter, we show that significant polarization can arise due to
dust scattering in the atmosphere of relatively warmer brown dwarfs where
dust grains have not been settled completely. In the presence of dust
 particulates of small size, the continuum flux in the near optical region
 could be polarized significantly while non-zero polarization can occur in
 the infra-red if the particle size is large. Hence observation of polarized
radiation from brown dwarfs could be a potential probe for the properties
of dust grains.

\section{NONSPHERICITY OF RAPIDLY ROTATING BROWN DWARFS AND
INTRINSIC POLARIZATION BY DUST SCATTERING}

Polarization of radiation due to scattering in stellar atmospheres or
envelopes has been considered by many authors. Numerous
polarimetric data and their interpretation have been
presented by Gehrels (1974). Since noticeable deviation from
sphericity are to be expected from rapidly rotating stars, much
attention has been paid to calculations of intrinsic polarization
of Be stars. Degree of polarization varying from 0.1 \% in the
visible region to 2\% in the far infra-red has been observed in Be stars
by  Rucinski (1970), Collins (1970), Harrington and Collins (1968), Haisch
and Cassinelli (1978). Extensive observational data for polarization
from cool stars are given by Serkowskii (1967, 1968). It has been
established by many observations that the intrinsic polarization of radiation
is a common behavior for most young stars (see for example Menard \&
Bastien 1992, Yudin \& Evans 1998 and references therein).

  Observation of non-zero polarization from unresolved objects
indicates non-sphericity of the stars. The non-sphericity of an
object, which leads to incomplete cancellation of the polarization
of radiation from different areas of the surface, may result from
several causes. The rotation of a stellar object imparts to it the shape
of an oblate ellipsoid.  This is evident in
the outer solar planets as well. The eccentricities of Jupiter,
Saturn and Uranus are  0.35, 0.43, 0.21 respectively at 1 bar pressure
level.  Tidal interaction with the companion in a binary
system imposes an ellipsoidal shape extended towards the companion.

 Spectroscopic studies by Basri (1999) indicate rapid rotation of
brown dwarfs along their axis. A general trend to higher velocities
as one looks to objects of lower luminosity is found in this study.
The brown dwarf Kelu 1 is found to be the fastest rotator having
the projected angular velocity $v\sin i$ as high as 80 km/s corresponding to
 an angular velocity of $1.16\times 10^{-3} s^{-1}$.
The surface gravity of brown dwarf varies from $10^5 cms^{-2}$
to $3\times 10^5 cms^{-2}$ implying a  mass range from 36 $M_J$ to
73 $M_J$ (Marley et al 1996).
As a consequence, a fast rotating brown dwarf will have significant
departure from sphericity.

  The maximum possible oblateness of a stable rotating fluid body
 having polytropic equation of state with different indices 
has been derived by Chandrasekhar (1933). The relation between the
eccentricity $e$ and the rotational velocity $\omega$ for a
rigidly rotating body (Maclaurine spheroid) with uniform density is given by :
 
\begin{equation}
\omega=\left[2\pi G \rho \left\{\frac{(1-e^2)^{1/2}}{e^3}(3-
2e^2)\sin^{-1}e-\frac{3}{e^2}(1-e^2)\right\}\right]^{1/2},
\end{equation}
where $\rho$ is the density of the fluid. Here the eccentricity is defined
as $e=(1-r_p^2/r_e^2)^{1/2}$ where $r_p$ and $r_e$ are the polar and the
equatorial radius respectively. The eccentricity `e' and
the oblateness `q' are related by the expression $e^2=1-(1-q)^2$.

Adopting the empirical relationship given in Marley et al. (1996)
for mass and radius of brown dwarf, the mean density of a brown
dwarf with $g=10^5 cms^{-2}$ and $T_{eff}=1500 K$ would be $51 gmcm^{-3}$.
For $\omega=1.16\times 10^{-3} s^{-1}$, the above formula yields
$e=0.48$. However, in reality the density is not uniform and
the above formula yields slightly higher value of e compared to
the same for non-uniform density distribution as can be 
verified by applying to the case of Jupiter, Saturn and Uranus. 
Nevertheless, the rapid rotation of brown dwarfs would certainly
impose deviation from sphericity in its shape and hence the disc
integrated polarization will not be cancelled out.

The dependence of polarization due to single scattering by grains on the
oblateness of an object has been discussed by Dolginov, Gnedin \&
Silant'ev (1995) and by Simmons (1982). Following Simmons (1982), the
analytical expression (first order approximation) for the degree of
polarization due to single scattering from a spheroid with
uniform density can be written as :
\begin{equation}
p=\lambda^{2}(R_1-R_2)n_0K_2(\frac{5}{96\pi})^{1/2}P^2_2\cos(i)\exp
(-2i\phi)F_{22}
\end{equation}
where $F_{22}$ depends on the scattering phase function, $R_1$ and $R_2$
are the outer and the inner equatorial axis lengths, $n_0$ is the
particle number density, $P^2_2$ is the associated Legendre
polynomial of order 2 and $\lambda$ is the wavelength. The expression for
$K_2$ depends on the density distribution and hence on the eccentricity
of the spheroid.  The values of $F_{22}$ for different values of $ x=2\pi a/\lambda$,
$a$ being the radius of the scattering particle, as well as the expression
for $K_2$ are given in Simmons
(1982). The first order approximation is valid when $x\leq 2$.
We have calculated the degree of polarization as a  function of eccentricity $e$
by considering Mie scattering. Figure~1 presents the degree of
polarization as a function of eccentricity  viewed edge on
 i.e., when  $i=\pi/2$ and $\phi=0$.  

\placefigure{figure1}

 Figure~1 shows that the qualitative feature of polarization
does not change with the change in the eccentricity. This is
found by Simmons (1982) as well. Polarization increases with
the increase in the eccentricity. For $x\geq 2$ the change
in the degree of polarization is small. For all the values of
$x$ the variation in polarization is not too high when $e$ varies
from 0.2 to 0.4. In our calculation we have taken $n_0a^2=100$.
The choice is arbitrary as the main purpose of Figure~1 is to show
the effect of eccentricity on the degree of polarization. In the
realistic case, the density must be non-uniform and the polarization
would occur due to multiple scattering. Therefore one has to solve
the equations for the transfer of polarized radiation with the
incorporation of Mie theory of scattering. 

\section{TRANSFER EQUATIONS FOR POLARIZED RADIATION}

The Stokes parameters representing a linearly polarized beam are given by
$I=I_l+I_r$ and $Q=I_l-I_r$ where $I_l$ and $I_r$ represent two mutually orthogonal
states of linear polarization. The degree of polarization is given by $p=Q/I$.

For a plane parallel atmosphere the equation of transfer can be written as
(Chandrasekhar 1960) :
\begin{equation}
\mu\frac{d{\bf I}(\tau,\mu)}{d\tau}={\bf I}(\tau,\mu)-\frac{\omega_0}{2}\int_{-1}^{1}
{\bf{P}(\mu,\mu')\bf{I}(\tau,\mu')d\mu'},
\end{equation}
where
$$\bf{I}(\tau,\mu)=\left(\begin{array}{c}I_l(\tau,\mu) \\ I_r(\tau,\mu)\end{array}\right),$$
$\mu$ is the cosine of the angle made by the ray to the normal, $\tau$ is the optical depth, $\omega_0$ is the albedo for single scattering and
$\bf{P}(\mu,\mu')$ is the azimuth-independent phase matrix. 

\section{THE ATMOSPHERE MODEL}

The values of the effective
temperature $T_{eff}$ and the surface gravity $g$ of brown dwarf
are constrained by the bolometric luminosity 
and the evolutionary sequence of the object (Saumon et al 1996). 
In the present letter, we have adopted a model atmosphere
with $T_{eff}=1500K$, $log(g)=5.0$ and $[M/H]=-0.3$ (K band). The 
temperature-pressure profile for model with $log(g)=5.0$ and $T_{eff}=1030 K$
has kindly been provided by M. Marley (private communication) and
the opacity data by D. Saumon. Using an iterative process for temperature
correction we have obtained the temperature profile for $T_{eff}=1500 K$. In
this process we have used the temperature profile for $T_{eff}=1030K$ as
the initial temperature profile. We have ignored any change in the
density profile with the increase in the temperature. We have incorporated
the dust opacity by using the Mie theory of scattering. Following Griffith,
Yelle and Marley (1998), and Marley \& Ackerman (2001) we have considered
 a lognormal spherical particle size distribution. The mean radius of the 
silicate grain (real part of the refractive index 1.65) has been considered
 as $ 0.1 \mu m$ and $1.0 \mu m$. Although it is difficult to infer
the grain sizes by direct observation, atmospheres of the outer solar
planets indicate that the mean radius
of the grains is expected to be as high as $10 \mu m$ to $100 \mu m$ 
Marley \& Ackerman 2001). However in the
present investigation we have considered smaller particles as the effect of
particle size on the degree of polarization can well be visualized without
going for larger particle size.

\section{RESULTS AND DISCUSSION}

We solve the transfer equations for the polarized radiation by using discretization
method. The numerical procedure is described in detail by Sengupta (1993). 

 In the absence of a magnetic field, a radiation field could be polarized only
by scattering. The extent of scattering is determined by the albedo $\omega_0$
and the angular distribution of the scattered photon is governed by the phase 
function. 
The albedo for single scattering is
defined as $\omega_0=(\sigma_R+\sigma_M)/\chi_{\nu}$ where $\sigma_R$ and
$\sigma_M$ are the cross-section for the Rayleigh and the Mie scattering respectively
and $\chi_{\nu}$ is the total extinction co-efficient. While the Rayleigh phase function
is symmetric in the sense that the amount of backward and forward scattering
are the same, the Mie phase function is asymmetric. The larger  the size of the dust
particulate, the greater the effect of the Mie scattering. 

When there are dust clouds at or above the photosphere, the albedo 
is high. Conversely, when clouds are absent, the albedo in the mostly absorbing
atmosphere is low (Burrows et al. 2001). As a consequence,
in the atmosphere of cool brown dwarf such as Gl229B, contribution of scattering
is negligible and so there will be no polarization. 
If the mean particle size is small, say $0.1 \mu m$,
 the albedo beyond near infra-red
is zero as the Mie scattering cross section goes over to that of the Rayleigh and
the scattering cross section of various molecules present in the atmosphere 
is negligibly small or zero. Therefore scattering polarization can arise only up to the optical
and near infra-red region where the contribution of scattering cross section to the
total extinction co-efficient is significant. 
It should be mentioned that  grains with small size cannot
contribute significantly to the opacities in the near-infrared where water
bands still dominate the spectra of brown dwarf from J to K band.

On the other hand if we consider larger particle size, then scattering
will contribute to the far infra-red region as well.

\placefigure{figure2}
The degree of polarization due to multiple scattering  is presented
 in figure~2. Here we notice that the
degree of polarization is almost zero  at $1.34 \mu m$ and onwards when the
mean particle size is $0.1 \mu m$.. Shortwards
of $1.3 \mu m$ the degree of polarization increases.  
This is because of the fact that scattering is negligible at wavelengths greater
than $1.3 \mu m$. As the wavelength decreases, the effect of dust scattering
increases. As a consequence the radiation field gets polarized due to
scattering and the degree of polarization increases with the decrease in
the wavelength. The degree of polarization always remains negative.

If we consider dust particles with larger mean radius, say $1 \mu m$, then
most of the flux get blocked in the optical region. As a result
the degree of polarization decreases significantly in the optical region.
However, when the size of the particle is comparable to the wavelength,
degree of polarization increases significantly. Figure~2 shows that
the degree of polarization decreases in the optical region if the
the particle size is comparatively large. In the infra-red and far infra-red,
degree of polarization increases if the grain  
size is increased. It's worth mentioning that multiple scattering
decreases the degree of polarization as compared to that with single
scattering.

\section{CONCLUSION}

  The important message that is conveyed in this letter is that the
continuum radiation from the atmosphere of relatively warm brown 
dwarfs where the dust particulates are not gravitationally
settled should be polarized and the polarized radiation could provide
a lot of information on the properties of dust particulates.
Since, no polarization will occur in the absence of dust grains,
observation of polarized radiation will also help in deciding
if a particular brown dwarf contains dust in its atmosphere.
Rapid rotation would impose non-sphericity in the shape of
brown dwarfs. As a result, a net non-zero polarization integrated
over the stellar disc should be obtained.
The results strongly suggest that the  polarimetric
observations posses a great diagnostic potential for
the understanding of the properties of dust particulates.   
Furthermore, if no polarization is observed from any particular
brown dwarf then that will indicate the absence of dust in
its atmosphere.

\acknowledgments
We thank M. Marley for kindly providing the temperature-pressure profile 
and  D. Saumon for the continuum opacity data. We are indebed to A. B.
Ravindran for many valuable discussions and to the referee for
constructive comments, suggestions and criticisms. Thanks are due to N. K. Rao,
P. Bhattacharya, H. C. Bhatt and T. P. Prabhu.

\clearpage

\figcaption[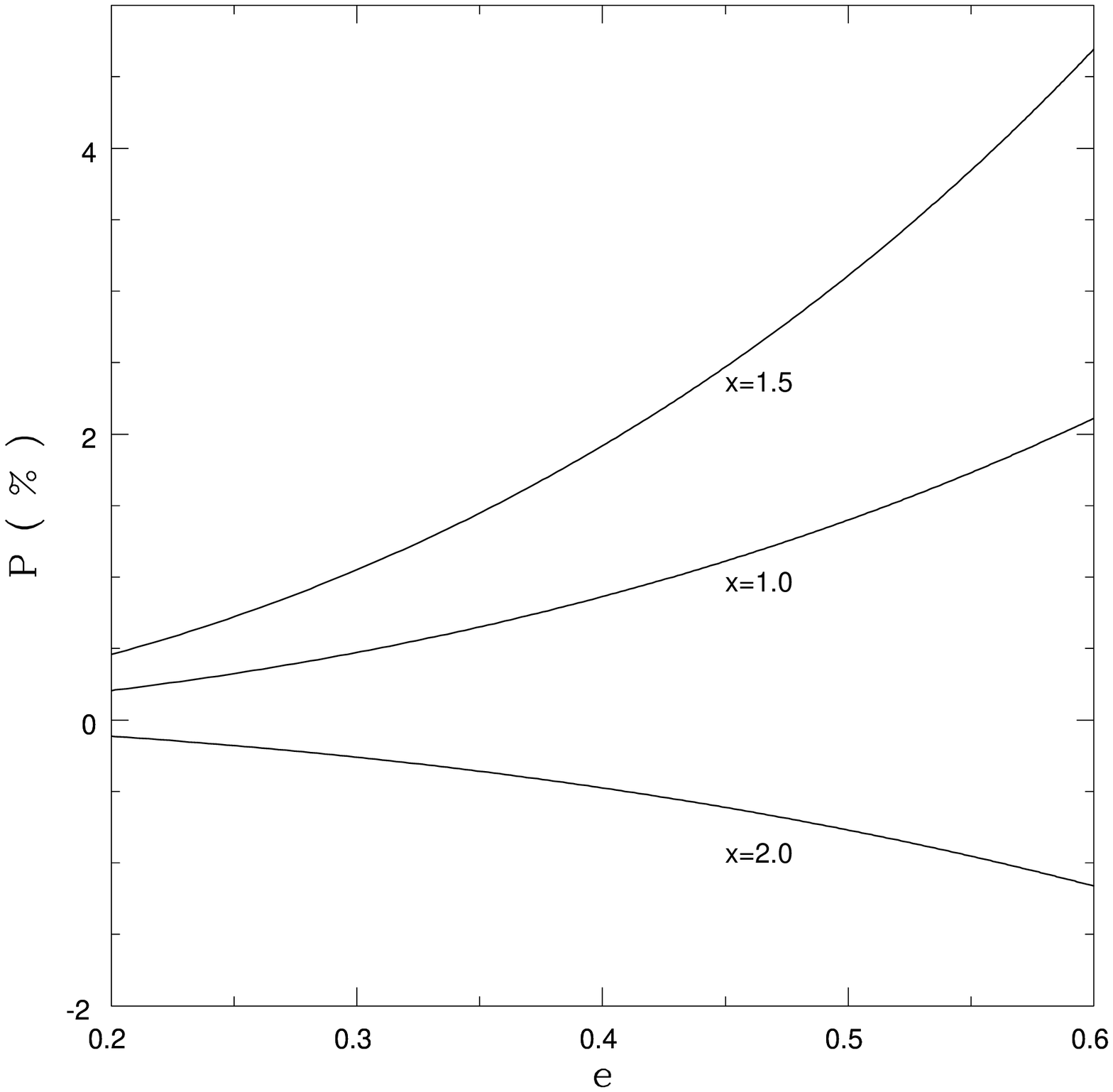]{Degree of polarization as a function of
eccentricity for different values of $x=2\pi a/\lambda$. Single scattering on a
spherical grain in an oblate spheroid with uniform density is considered.
\label{figure1}}

\figcaption[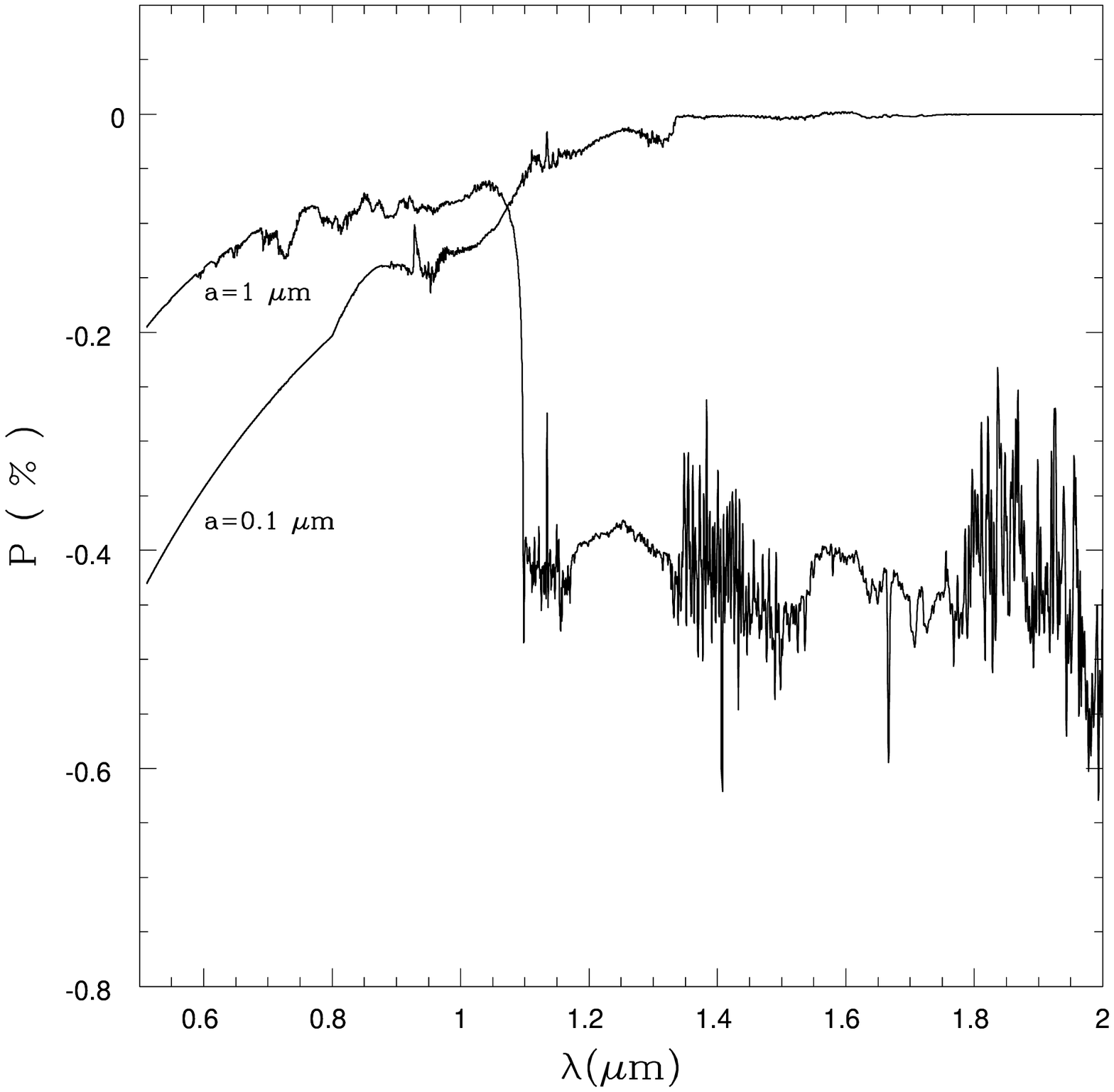]{Degree of polarization by multiple scattering.
\label{figure2}}

\end{document}